\documentclass[]{WileyASNA-v1}

\usepackage{amssymb}
\usepackage{graphicx}
\usepackage{times}
\usepackage{amsmath,siunitx}
\usepackage[utf8]{inputenc}
\usepackage[T1]{fontenc}
\usepackage{hyperref}
\usepackage{ulem}

\articletype{Article Type}

\received{2023}
\revised{2023}
\accepted{2023}

\newcommand{\white}[1]{\textcolor{white}{#1}}

\begin{document}

\title{The deep eclipses of RW\,Aur revisited by long-term photometric and spectroscopic monitoring\protect\thanks{Based on observations obtained with telescopes of the University Observatory Jena, which is operated by the Astrophysical Institute of the Friedrich-Schiller-University.}}

\author[1]{Oliver Lux}

\author[1]{Markus Mugrauer}

\author[1]{Richard Bischoff}

\authormark{Lux, Mugrauer \& Bischoff}

\address[1]{Astrophysikalisches Institut und Universit\"{a}ts-Sternwarte Jena}

\corres{Oliver Lux, Astrophysikalisches Institut und Universit\"{a}ts-Sternwarte Jena, Schillerg\"{a}{\ss}chen 2, D-07745 Jena, Germany.\newline \email{oliverlux@gmx.de}}

\abstract{RW\,Aurigae is a young stellar system containing a classical T Tauri star as the primary component. It shows deep, irregular dimmings, first detected in 2010. At the University Observatory Jena, we carried out optical follow-up observations. We performed multiband (BVRI) photometry of the system with the \textit{Cassegrain-Teleskop-Kamera II} and the \textit{Schmidt-Teleskop-Kamera} between September 2016 and April 2019, as well as spectroscopy, using the \textit{Fibre Linked \'ECHelle Astronomical Spectrograph} between September 2016 and April 2018. We present the apparent photometry of RW\,Aur, which is consistent with and complementary to photometric data from the \textit{American Association of Variable Star Observers}. The V-band magnitude of RW\,Aur changed by up to three magnitudes during the timespan of our monitoring campaign. For the observing epochs 2016/2017 and 2017/2018 we report a decreasing brightness, while in the epoch 2018/2019 the system remained in a relatively constant bright state. In the color-magnitude diagram, we see that RW\,Aur lies close to a track for grey extinction. The spectra show a decreasing equivalent width of the $\rm{H}\alpha$ emission line for decreasing brightness, whereas the equivalent width of the $[\rm{O}I]$ line increases, indicating an increased outflow activity during the obscuration. Both gives further evidence for the favored theory that the obscurations are caused by a hot dusty wind emerging from the inner disk.}


\maketitle

\section{Introduction} \label{sec1}

\begin{figure*}[h!]
\resizebox{\hsize}{!}{\includegraphics{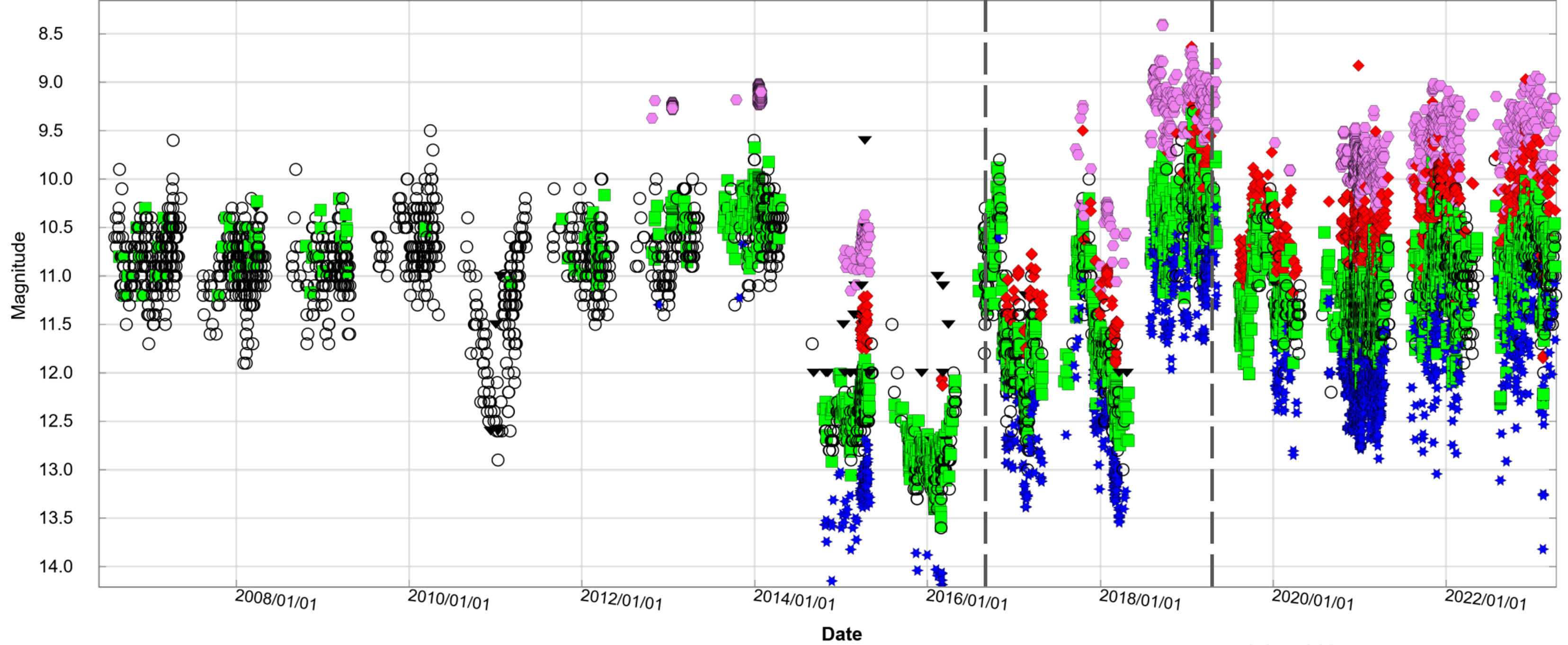}}\caption{The light curve of RW\,Aur in optical filters between 2006/04/01 and 2023/04/07, taken from AAVSO. Blue stars give B-band, green squares V-band, red diamonds R-band and purple hexagons I-band magnitudes. Black triangles indicate upper limits. The dark-grey dashed lines mark the beginning and end of our photometric monitoring campaign, respectively.}
\label{aavso}
\end{figure*}

T Tauri stars (TTS) are low-mass pre-main sequence stars with masses up to about 2\,$\rm{M}_\odot$ \citep{Joy1945, Shenavrin2015}, which are usually surrounded by circumstellar disks made of gas and dust. They show significant variability on different timescales, which is mainly caused by strong surface magnetic fields leading to cool spots, as well as mass accretion leading to hot spots \citep{Herbst1994}. TTS which show both phenomena are usually referred to as classical TTS, while weak-line TTS lack the accretion activity.

\begin{table}
\caption{Stellar parameters of RW\,Aur\,A.}
\label{stellardata}
\centering
\begin{tabular}{ccc}
\hline
Parameter & Value & Reference \\
\hline
RA (J2000)                    & 05\,h 07\,m 49.6\,s                                         & 1   \\
Dec (J2000)                   & $+$30\,d 24\,m 05\,s                                        & 1   \\
$\mu_{\text{RA}}$             & $4.906\pm0.329$\,mas/yr                                     & 1   \\
$\mu_{\text{Dec}}$            & $-24.578\pm0.255$\,mas/yr                                   & 1   \\
$\varpi$                      & $5.4541\pm0.3324$\,mas                                      & 1   \\
Distance                      & $185\pm10$\,pc                                              & 2   \\
Mass                          & $1.34\pm0.18\,\rm{M}_{\odot}$                               & 3   \\
SpT                           & $\rm{K}2\pm2$                                         & 4   \\
$T_{\rm eff}$                 & $5040_{-420}^{+240}$\,K                                     & 5   \\
log $L[\rm{L}_{\odot}]$       & $0.23\pm0.14$                                               & 3   \\
X-ray count rate             & $0.020\pm0.009\,\rm{s}^{-1}$                              & 6 \\
Accretion rate                & $1.5\text{ - }2.0\times10^{-8}\,\rm{M}_{\odot}/\rm{yr}$     & 7,8 \\
Age                           & $3\pm1$\,Myr & 9 \\
\hline
\end{tabular}

\vspace{3mm}

(1) \citet{Gaia2022}, (2) \citet{Bailer-Jones2021}, (3) \citet{White2001}, (4) \citet{White2004}, (5) \citet{Pecaut2013}, (6) \citet[ROSAT HRI]{Koenig2001}, (7) \citet{Ingleby2013}, (8) \citet{Koutoulaki2019}, (9) \citet{Dodin2020}

\end{table}

\begin{table*}[h]
\caption{Observing log for our photometric monitoring of RW\,Aur. We list for each observing epoch, the covered span of time, the used instrument, the number of nights when observations were carried out ($N_{\text{Obs}}$), the detector integration times ($DIT$ in units of seconds), and the diameter in pixel ($\diameter$) of the used apertures in the B-, V-, R-, and I-band.}
\label{exptimes}

\begin{tabular}{cccc|cc|cc|cc|cc}
\hline
Epoch    & Dates                   & Instrument   & $N_{\text{Obs}}$ & $DIT(\rm{B})$ & $\diameter$ & $DIT(\rm{V})$ & $\diameter$ & $DIT(\rm{R})$ & $\diameter$ & $DIT(\rm{I})$ & $\diameter$  \\
\midrule
I        & 2016/09/07 - 2017/04/20 & CTK-II       & 19               & 90            & 5           & 30            & 5            & 30           & 5           & 30            & 5            \\
         \midrule
IIa      & 2017/09/05 - 2017/09/19 & STK          & 3                & 30            & 4           & 10            & 4            & 10           & 4           & 10            & 5            \\
         & 2017/10/15 - 2018/01/15 & STK          & 6                & 30            & 4           & 5.0           & 4            & 5.0            & 4           & 5.0           & 5            \\
IIb      & 2018/01/30 - 2018/03/29 & CTK-II       & 10               & 90            & 4           & 30            & 4            & 30           & 4           & 30            & 4            \\
\midrule
III      & 2018/08/16 - 2018/11/18 & STK          & 30               & 30            & 6           & 5.0           & 5            & 5.0          & 5           & 5.0           & 5            \\
         & 2019/01/18 - 2019/04/21 & STK          & 29               & 10            & 6           & 2.5           & 5            & 2.5          & 5           & 2.5           & 5            \\
\hline
\end{tabular}
\end{table*}

RW\,Aurigae (RW\,Aur) was one of the first stars to be identified as a TTS \citep{Joy1945}. The star is a target for photometric observations for more than 50 years and long known to be variable. Its basic stellar parameters are summarized in\linebreak Table\,\ref{stellardata}\hspace{-1.5mm}. RW\,Aur is located in the Taurus-Auriga star forming region at a distance of about $185\pm10$\,pc \citep{Bailer-Jones2021}, consistent with the \textit{Gaia} DR3 parallax distance of $183\pm11$\,pc \citep{Gaia2022}. Today it is known to be a multiple system, where the primary star, RW\,Aur\,A, is a classical TTS with spectral type $\rm{K}2\pm2$ \citep{White2004} and a mass of $1.34\pm0.18\,\rm{M}_{\odot}$ \citep{White2001}. The secondary star in the system, RW\,Aur\,B, is separated from its primary by about 1.49\,arcsec \citep[based on Gaia DR3 astrometry, ][]{Gaia2022}, which corresponds to a projected separation of about 270\,au. RW\,Aur\,B is a weak-line TTS, which has a spectral type $\rm{K}6\pm1$ \citep{White2004} and a mass of about 0.8\,$\rm{M}_{\odot}$ \citep{Shenavrin2015}. It was investigated in detail and classified as an UX\,Ori type variable star by \citet{Dodin2020}. \citet{Ghez1993} reported the discovery of a tertiary component, RW\,Aur\,C, which is probably a substellar companion with a mass of less than 0.045\,$\rm{M}_{\odot}$ \citep{Ghez1997}. It exhibits an angular separation to RW\,Aur\,B of 0.12\,arcsec (or 20\,au of projected separation). Hence, RW\,Aur is a hierarchical triple system.

Figure\,\ref{aavso}\hspace{-1.5mm}shows the light curve of RW\,Aur between April 2006 and April 2023, created with the light curve generator from the \textit{American Association of Variable Star Observers} (AAVSO\footnote{Online available at: \url{https://www.aavso.org/}}). It can be seen that the system usually has \mbox{$V=10.5\pm1$\,mag}, where the photometric variations happen on a time-scale of days \citep{Shenavrin2015}. RW Aur is very active, showing strong spectral signatures of accretion and winds \citep{Petrov2001,Alencar2005,Bozhinova2016}.

In 2010, the system started to show an unusual behavior, when it faded by about 2\,mag compared to its regular state. This dimming lasted for about half a year and was followed by an even longer and deeper dimming (about 3\,mag) between 2014 and 2016. After that, until 2020, RW\,Aur showed strong, unpredictable variability with amplitudes almost as in the second eclipse, but returning to the bright state at least once per year. Since 2020, the brightness becomes more stable, however on a slightly fainter level compared to the original brightness and a local peak in the observing epoch 2018/2019. The physical nature of these obscurations, and if they are all caused by the same occulting body, is not yet fully understood.

\citet{Cabrit2006} discovered a trailing arm around RW\,Aur\,A, which was later explained to be produced during the periastron passage of RW\,Aur\,B \citep{Dai2015}, which happened about 500 years ago \citep{Koutoulaki2019}. According to \citet{Berdnikov2017}, this encounter also caused a rearrangement of the inner disk and also triggered the jet, which is observed to emerge from RW\,Aur\,A \citep{Hirth1994}. \citet{Rodriguez2013} suggested that the tidal arm caused the occultation observed between 2010 and 2011.

However, \citet{Shenavrin2015} discovered that RW\,Aur shows an IR excess during the eclipses, indicating hot dust in the line of sight \citep[with a temperature of about 500 to 700\,K,][]{Koutoulaki2019}. This would not be expected for an occultation by a tidally induced spiral arm, which would be too distant to the star. Due to the inclination of the outer circumstellar disk \citep[$i=45\text{ - }60\,^{\circ}$,][]{Facchini2016}, the occultations cannot be caused by the outer parts of the disk. \citet{Petrov2015} suggested that a wind emanating from the star or the inner disk could carry large (about 1\,$\mu$m) dust grains into the line of sight, which would account for the observed grey extinction (e.g. \citealt{Antipin2015}). The photometric and spectroscopic observations by \citet{Bozhinova2016}, focusing on the second deep eclipse, supported this scenario of a hot, dusty wind.

\citet{Facchini2016} suggested that the occultations might be caused by a misaligned or warped inner disk, resulting either from the tidal encounter with RW\,Aur\,B or from a yet unknown (sub-)stellar companion around RW\,Aur\,A. \citet{Facchini2016} and \citet{Koutoulaki2019} found unchanged signs of accretion during the dimming, further indicating that the occulting dust layer lies close to the primary star (within about 1\,au) and only obscures the inner 0.05 to 0.1\,au of the system \citep{Koutoulaki2019}.

A further mechanism to explain the perturbations of the inner disk was presented by \citet{Guenther2018}, who performed X-ray observations and detected an increase of the coronal iron abundance of more than an order of magnitude during the dim state. They suggested that a breakup of planetesimals or a terrestrial planet could have enriched the inner disk with iron-rich dust, which would then be accreted by RW\,Aur\,A, increasing the stellar iron abundance. This scenario is further supported by \citet{Lisse2022}, who presented evidence for planetesimal formation, migration and destruction from IR spectroscopy.

\section{Observations and data reduction} \label{sec2}

We started the photometric and spectroscopic monitoring campaign in August 2016 after our discovery of the re-brightening of RW\,Aur after the second deep eclipse \citep{Scholz2016}. Observations of RW\,Aur were taken between 31 August 2016 and 21 April 2019 with instruments operated at the University Observatory Jena, which is located near the small village Gro{\ss}schwabhausen about 10\,km west of Jena \citep{Pfau1984}. The photometry of the target was monitored in 58 nights with the \textit{Cassegrain-Teleskop-Kamera II} \citep[CTK-II from hereon,][]{Mugrauer2016}, and in 39 nights with the \textit{Schmidt-Teleskop-Kamera} \citep[STK from hereon,][]{Mugrauer2010}. In addition, we could obtain spectroscopic observations of RW\,Aur in 11 nights with the \textit{Fibre Linked \'ECHelle Astronomical Spectrograph} \citep[FLECHAS from hereon,][]{Mugrauer2014}. The photometric and spectroscopic observations, including the data reduction, are explained in detail in the following two subsections.

\subsection{Photometric observations}

In the course of our monitoring campaign, starting on 07 September 2016, we observed RW\,Aur in total in 97 nights with the CTK-II and STK, and obtained unresolved photometry of the target (i.e. RW\,Aur\,A and B are not individually detected) in the B-, V-, R-, and I-band. Our observing campaign of RW\,Aur can be divided into three epochs, each ranging from late summer to early spring, corresponding to the visibility of the star. Table\,\ref{exptimes}\hspace{-1.5mm} gives the used instruments for the different epochs as well as the detector integration time ($DIT$) used for the different filters. Three exposures per filter were taken in each night, and the $DIT$ was adjusted dependent on the used instrument, the seeing conditions, and the brightness of the target.

The standard reduction of all data, i.e. dark subtraction and flat-fielding, was performed with \textit{ESO Eclipse} \citep{Devillard2001}. For observations with the CTK-II, for the flat-fielding we used skyflats, which were always taken in the morning or evening twilight in the same night as the observations of the target if possible given the weather conditions, otherwise in the next consecutive night. For the flat-fielding of the STK data we used domeflats, which were always taken at the end of each observing night. Dark frames for all needed integration times were taken in each night for both instruments. The fully reduced images were then used for the photometric analysis, which is described in Section~\ref{sec3}.

\subsection{Spectroscopic observations}

For our spectroscopic monitoring observations of RW\,Aur we used FLECHAS in its $2\times2$-binning mode, which yields a resolving power of about $R=7300$ over the full spectral range between 3900 and 8100\,\AA, covered by the instrument. The follow-up spectroscopy of the system was carried out between 31 August 2016 and 10 April 2018. In each observing night, always three spectra of the target were taken with $DIT$ of 900\,s, 1200\,s or 1800\,s, chosen dependent on the brightness of the star, to guarantee that spectra with a sufficiently high signal-to-noise ratio ($SNR$) could be obtained ($SNR>40$).

\begin{table}[h!]
\caption{Observing log for our FLECHAS follow-up spectroscopy of RW\,Aur. We list for each observing epoch, the Gregorian and Julian date ($JD$), the used detector integration time ($DIT$), the signal-to-noise ratio ($SNR$) at the peak of the $\rm{H}\alpha$ emission line in the fully-reduced spectrum, and the mean airmass $X$ of the target during the observations.}
\label{Spec_exptimes}
\resizebox{\hsize}{!}{\begin{tabular}{cccccc}
\toprule
Epoch   & Date        & $JD$        & $DIT [s]$ & $SNR$ & $X$\\
\midrule
I       &  2016/08/31 & 2457631.574 & 900       & 191   & 1.4 \\
        &  2016/09/02 & 2457633.557 & 1200      & 167   & 1.5 \\
        &  2016/09/13 & 2457644.569 & 1200      & 225   & 1.3 \\
        &  2016/09/15 & 2457646.581 & 1200      & 251   & 1.2 \\
        &  2016/09/21 & 2457653.499 & 1200      & 239   & 1.5 \\
        &  2016/09/29 & 2457661.458 & 1200      & 290   & 1.7 \\
        &  2017/02/15 & 2457800.391 & 1800      & 129   & 1.3 \\
\midrule
IIb     &  2018/02/23 & 2458173.371 & 900       & 50    & 1.3 \\
        &  2018/02/25 & 2458175.306 & 1800      & 85    & 1.1 \\
        &  2018/03/01 & 2458179.429 & 1800      & 61    & 1.9 \\
        &  2018/04/10 & 2458219.339 & 900       & 41    & 2.0 \\
\bottomrule
\end{tabular}}
\end{table}

In each observing night, directly before the spectroscopic observation of RW\,Aur, we recorded three spectra of a tungsten lamp for flat-fielding, as well as three spectra of a Thorium-Argon (ThAr) lamp for wavelength calibration, each with an exposure time of 1.5\,s. The dark-frames for the flat-fielding and wavelength calibration were taken in the beginning of each observing night, while the dark-frames with longer integration times, needed for the science spectra, were taken in longer time intervals (up to a few weeks). The data reduction was performed with the FLECHAS data reduction pipeline \citep{Mugrauer2014}. The individual reduction steps covered by the pipeline are dark- and bias-subtraction, flat-fielding, extraction of the individual spectral orders, wavelength calibration and normalization. The long-term stability of the instrument has been demonstrated by \citet{Bischoff2017} and \citet{Heyne2020}. Details of our follow-up spectroscopy of RW\,Aur are summarized in Table\,\ref{Spec_exptimes}\hspace{-1.5mm}.

\section{Photometric analysis} \label{sec3}

The photometric analysis of RW\,Aur was carried out utilizing aperture photometry. At first the photometric variability of all detected stars in the images was determined with \textit{Muniwin} \citep{Hroch2014}, and photometric reference stars were chosen based on the following criteria:

\begin{itemize}
\item exhibit a minimal photometric variability throughout the whole photometric monitoring campaign (on the 1\% level or better), identified by an automatic comparison of the photometric variability of the instrumental magnitudes of all stars, detected in the field of view of the used instruments
\item be bright (within $\sim$4\,mag of RW\,Aur during its bright state) but faint enough to avoid saturation of the used instruments during the whole photometric monitoring campaign
\item be located within 10\,arcmin around RW\,Aur, to minimize airmass differences to the target, and to guarantee their detection with all used instruments in all observing epochs
\end{itemize}

\begin{figure}
\resizebox{\hsize}{!}{\includegraphics{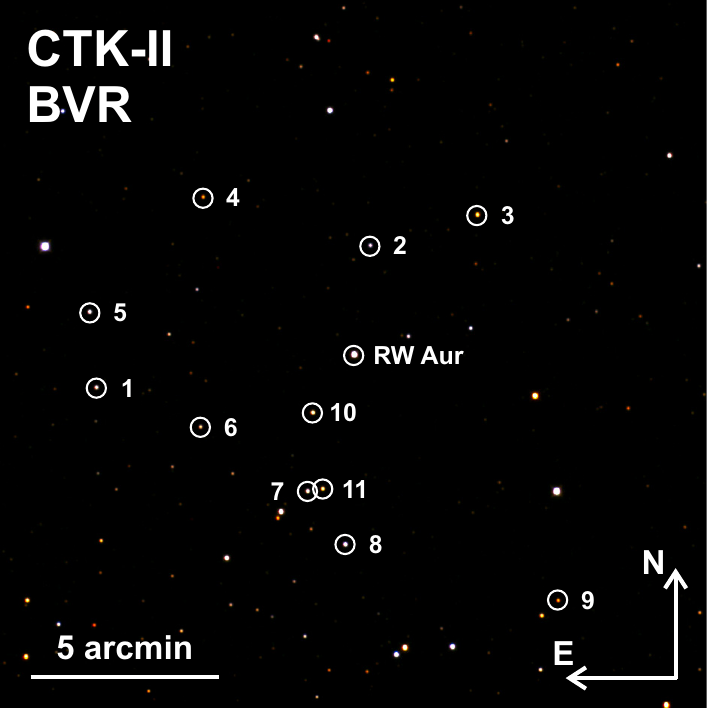}}
\caption{RGB-color-composite image of RW\,Aur, taken with the CTK-II on 15 September 2016. The image is made of R-, V-, and B-band images, with total integration times of 90, 90, and 270\,s, respectively. RW\,Aur is located in the center of the image. The selected photometric reference stars are indicated and marked with white circles.}
\label{refstars}
\end{figure}

Figure\,\ref{refstars}\hspace{-1.5mm} shows a CTK-II RGB-color-composite image of RW\,Aur and the selected photometric reference stars, made of images taken in the R-, V-, and B-band.

We then checked the quality of all images and discarded those where RW Aur is saturated or has a low $SNR$. In total, 1116 images (97.1\,\% of all) from 97 nights have sufficient quality and are therefore included in the photometric analysis of RW\,Aur. In detail, photometric data of the target could be obtained in 89 nights in the B-band, in 95 nights in the V-band, in 96 nights in the R-band, and in 93 nights in the I-band, respectively.

In order to determine the apparent magnitude of RW\,Aur in the standard (Johnson-Cousins) photometric system, the magnitudes of the reference stars in this system are needed, which were taken from the URAT1 catalog \citep{Zacharias2015}. This catalog lists the Johnson B- and V-band magnitudes, as well as the g-, r- and i-band magnitudes from the photometric system of the Sloan Digital Sky Survey \citep[SDSS]{York2000}. We convert the SDSS magnitudes of the reference stars to their R-, and I-band magnitudes in the Johnson-Cousins system, using the empirical color-transformation-relations: $\rm{R} = \rm{V} - (\rm{g} - \rm{r} + 0.139) / 1.646$ and $\rm{I} = \rm{R} - (\rm{r} - \rm{i} + 0.236) / 1.007$, determined by \citet{Jordi2006}. The apparent magnitudes of the reference stars for all bands are summarized in Table\,\ref{reflitmag}\hspace{-1.5mm}.

We then measured the instrumental magnitudes of RW\,Aur and of the reference stars in all images. Thereby, dependent on the seeing conditions, the focussing of the instrument during the individual observing epochs, and the used filters, we varied the aperture size between 4 and 6 pixels, corresponding to about twice the full width at half maximum (FWHM) of the seeing disks of the stars detected in the images. The inner and outer diameter of the background annulus was kept constant at 10 and 20 pixels, respectively. The used apertures for all observing epochs and filters are summarized in Table\,\ref{exptimes}\hspace{-1.5mm}.

For each object we determine the average and standard deviation of its instrumental magnitudes in each observing night, and calculate the differential magnitude between RW\,Aur and the photometric reference stars in all filters. With the given magnitudes of the reference stars in the photometric standard system this yields the apparent B-, V-, R-, and I-band magnitudes of RW\,Aur for all observing nights. The determined apparent magnitudes of RW\,Aur for all photometric bands over the whole time span of our photometric monitoring campaign are illustrated as light curves in Figure\,\ref{lightcurve}\hspace{-1.5mm}and summarized in the Appendix in Table\,\ref{photodata}\hspace{-1.5mm}. The photometric uncertainties are below 15\,mmag and hence not shown in this plot, as they are smaller than the size of the used data-points. The apparent photometry of RW\,Aur, measured in our campaign, is consistent and complementary with the AAVSO photometry of the star available for the same range of time.

\begin{figure}
\resizebox{\hsize}{!}{\includegraphics{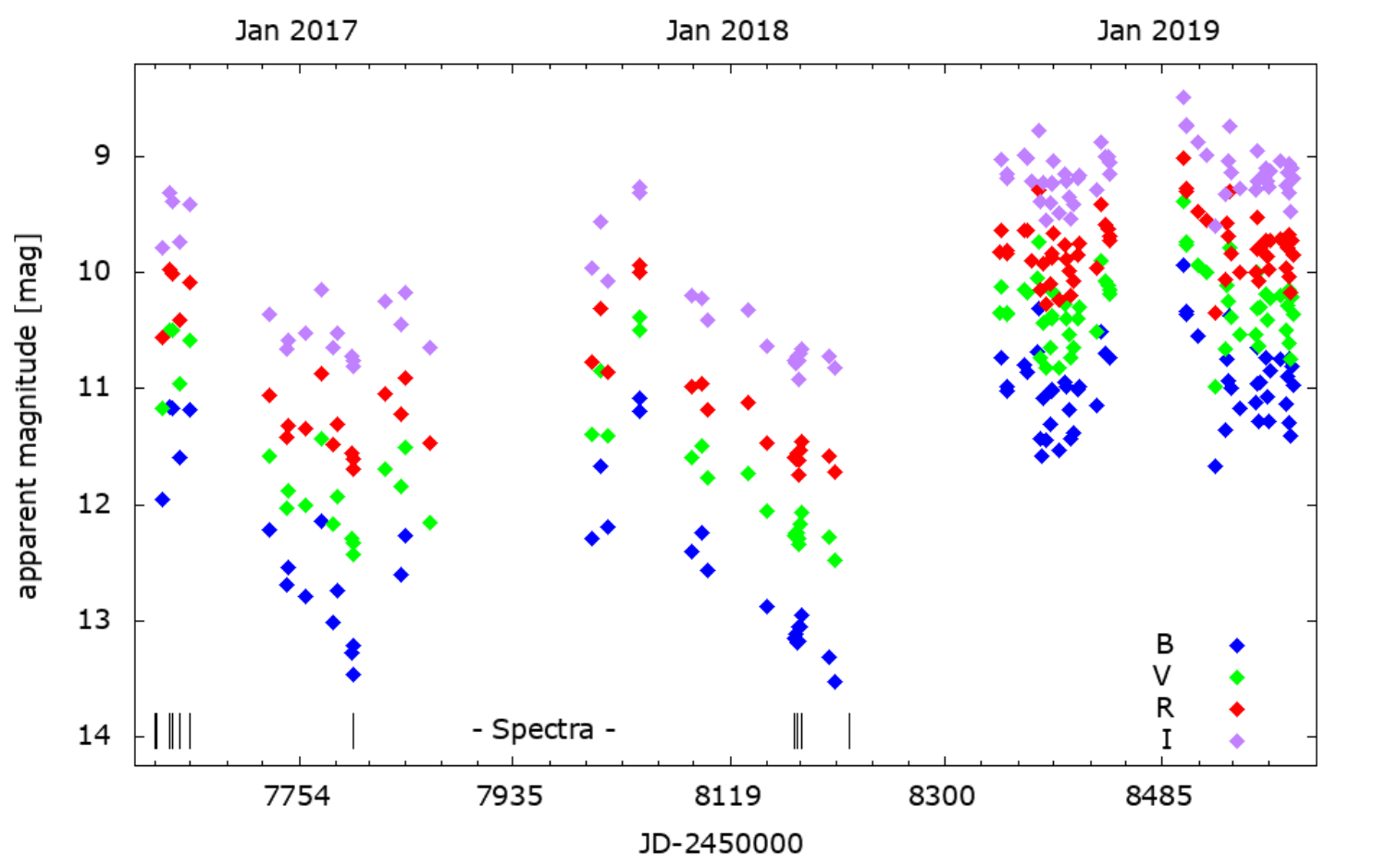}}
\caption{The light curve of RW\,Aur for the whole span of time covered by our photometric monitoring project. The diamonds show the apparent B-, V-, R-, and I-band magnitudes of the star, and the black lines indicate the points in time, when FLECHAS spectra of the star were taken.}
\label{lightcurve}
\end{figure}

\begin{figure*}
\resizebox{\hsize}{!}{\includegraphics{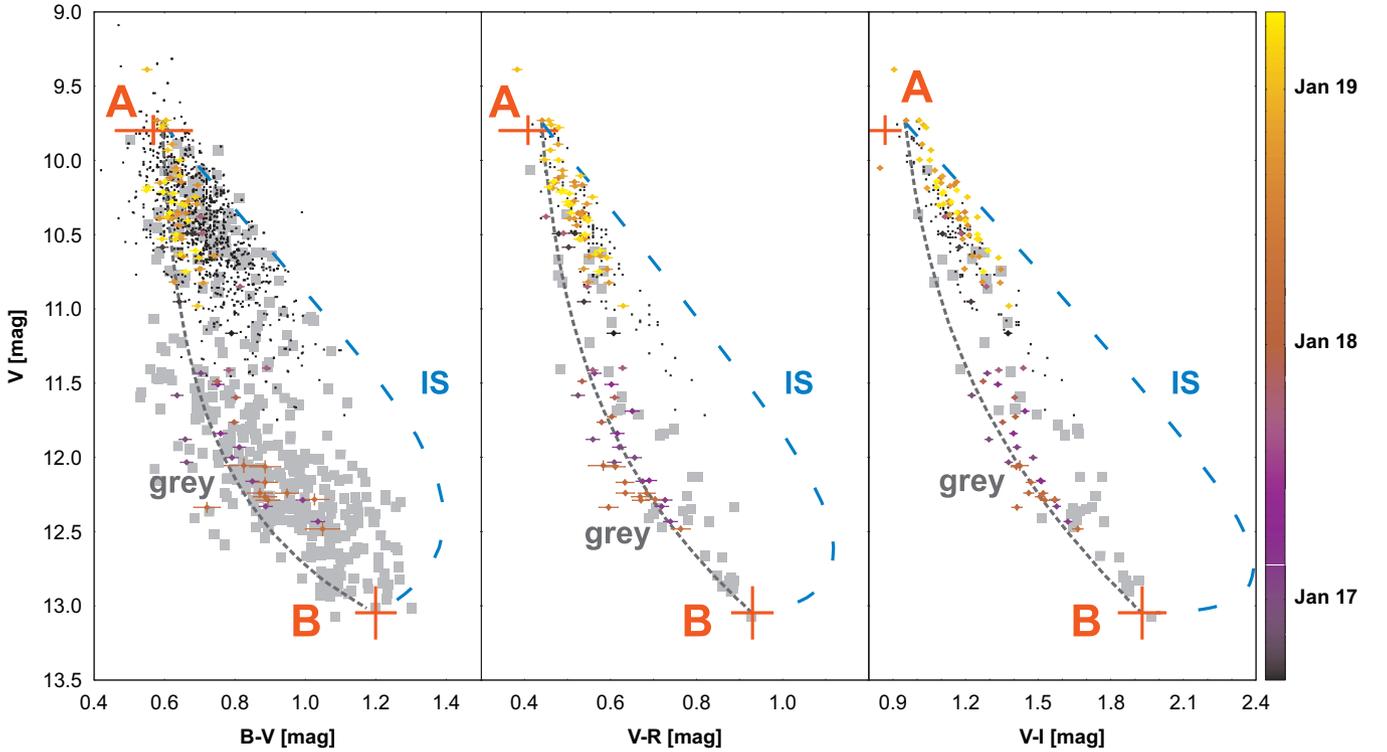}}
\caption{Color-magnitude diagrams of RW\,Aur. The abscissas show the $(\rm{B}-\rm{V})$, $(\rm{V}-\rm{R})$, and $(\rm{V}-\rm{I)}$ colors of the stellar system, while its apparent V-band magnitude is plotted along the ordinates. The observation date is color-coded for our observations and the photometric uncertainties are indicated with error bars. Black and grey squares show data presented by \citet{Dodin2019} and correspond to observations taken before and after September 2010, respectively. The typical photometric errors for these observations are 0.2\,mag in the B-, and 0.1\,mag in the V-, R-, and I-band, respectively. The wide-dashed blue line in all diagrams indicates the color-magnitude-track of the RW\,Aur system for interstellar extinction according to the standard extinction law ($R_{\rm V} = 3.1$),  while the narrow-dashed grey line shows the track for grey extinction. A and B mark the brightness and color of RW\,Aur\,A during its bright state, and RW\,Aur\,B.}
\label{cmds}
\end{figure*}
\begin{table*}
\caption{The apparent magnitudes of the selected photometric reference stars, numbered as in Figure\,\ref{refstars}\hspace{-1.5mm}. The typical photometric uncertainty of the reference stars in the different bands is 0.03\,mag.}
\label{reflitmag}
\centering
\begin{tabular}{ccccccccc}
\hline
Ref. star \# & URAT1 ID   & B [mag] & V [mag] & R [mag] & I [mag] & g [mag] & r [mag] &  i [mag]\\
\hline
1          & 602-067264 & 13.217  & 12.571  & 12.202  & 11.737  & 12.832  & 12.364  & 12.132  \\
2          & 603-068385 & 13.132  & 12.707  & 12.492  & 12.167  & 12.852  & 12.637  & 12.546  \\
3          & 603-068220 & 13.744  & 12.316  & 11.510  & 10.769  & 13.009  & 11.821  & 11.311  \\
4          & 603-068622 & 14.876  & 13.348  & 12.487  & 11.709  & 14.101  & 12.823  & 12.275  \\
5          & 603-068798 & 12.936  & 12.361  & 12.043  & 11.635  & 12.589  & 12.205  & 12.030  \\
6          & 602-067100 & 13.895  & 13.090  & 12.639  & 12.146  & 13.441  & 12.837  & 12.577  \\
7          & 602-066917 & 13.070  & 12.353  & 11.948  & 11.471  & 12.657  & 12.130  & 11.885  \\
8          & 602-066877 & 12.571  & 12.046  & 11.751  & 11.369  & 12.252  & 11.905  & 11.757  \\
9          & 602-066582 & 15.164  & 13.519  & 12.594  & 11.684  & 14.319  & 12.936  & 12.255  \\
10         & 602-066915 & 13.270  & 12.223  & 11.601  & 11.012  & 12.723  & 11.839  & 11.481  \\
11         & 602-066898 & 13.812  & 12.535  & 11.824  & 11.114  & 13.133  & 12.101  & 11.622  \\
\hline
\end{tabular}\vfill
\end{table*}

The determined apparent photometry of RW\,Aur in all photometric bands can be used to study the color evolution of the target, dependent on its apparent magnitude, between its faint and bright state, which gives insights on the nature of the obscuration. Color-magnitude-diagrams of RW\,Aur are shown in Figure\,\ref{cmds}\hspace{-1.5mm}. The color-coded dots in these diagrams are the data from our photometric monitoring campaign. These can be compared to data from \cite{Dodin2019}, who presented own photometric observations accompanied by data from \citet{Herbst1994}, \citet{Grankin2007} and the AAVSO. The black and grey squares in the diagrams represent observations from before and after September 2010, when the first deep dimming in the light curve of RW\,Aur was detected. The evolution of the color of RW\,Aur can be compared to interstellar extinction \citep{Savage1979}, indicated by the blue line, which causes rather red colors, and grey extinction (grey line), which is connected to large dust grains along the line of sight, resulting in rather equal obscuration in the different photometric bands. The position of RW\,Aur A during its bright state was constructed by \citet{Dodin2019} such that the curves for grey- and interstellar extinction envelope most observations. The position of RW\,Aur\,B was obtained from their resolved photometric observations of the companion.

Our measurements are consistent with the ones presented by \citet{Dodin2019}. The color and magnitude of the system is close to the color and magnitude of RW\,Aur\,A during the bright state and close to the color and magnitude of RW\,Aur\,B during the faint state. While during bright state the colors are in the middle between the tracks for grey and interstellar extinction, we see a clear tendency towards grey extinction when the system becomes fainter. This gives good evidence that the obscuration is caused by large dust grains along the line of sight.

\citet{Koutoulaki2019} obtained spectra of RW Aur A with X-shooter \citep{Vernet2011}, one during the faint state in March 2015, and one during the bright state in September 2016. By dividing the two spectra, they can compare the evolution of the optical depth over wavelength with the track for interstellar extinction \citep{Cardelli1989} as well as models for dust extinction with different grain sizes and amounts of scattering. They find that the extinction is between interstellar and grey for the wavelength range covered by their observations (0.3\,-\,2.4\,$\mu$m). This also applies to the part of their spectra that can be compared to our BVRI photometry (0.4\,-\,0.9\,$\mu$m). Despite this difference to our findings in the faint state, where the extinction is clearly grey, both studies get to the conclusion that the obscuring screen mainly consists of large dust grains. \citet{Koutoulaki2019} obtain maximum grain sizes of 15\,-\,150\,$\mu$m, depending on the amount of scattering and the grain types. Also note that the comparison has to be taken with caution, because of the different methods, and because \citet{Koutoulaki2019} obtained the spectrum for the faint state in an earlier epoch, when the dimming was stronger.

\section{Spectral analysis} \label{sec4}

\begin{figure*}
\resizebox{\hsize}{!}{\includegraphics[height=5cm]{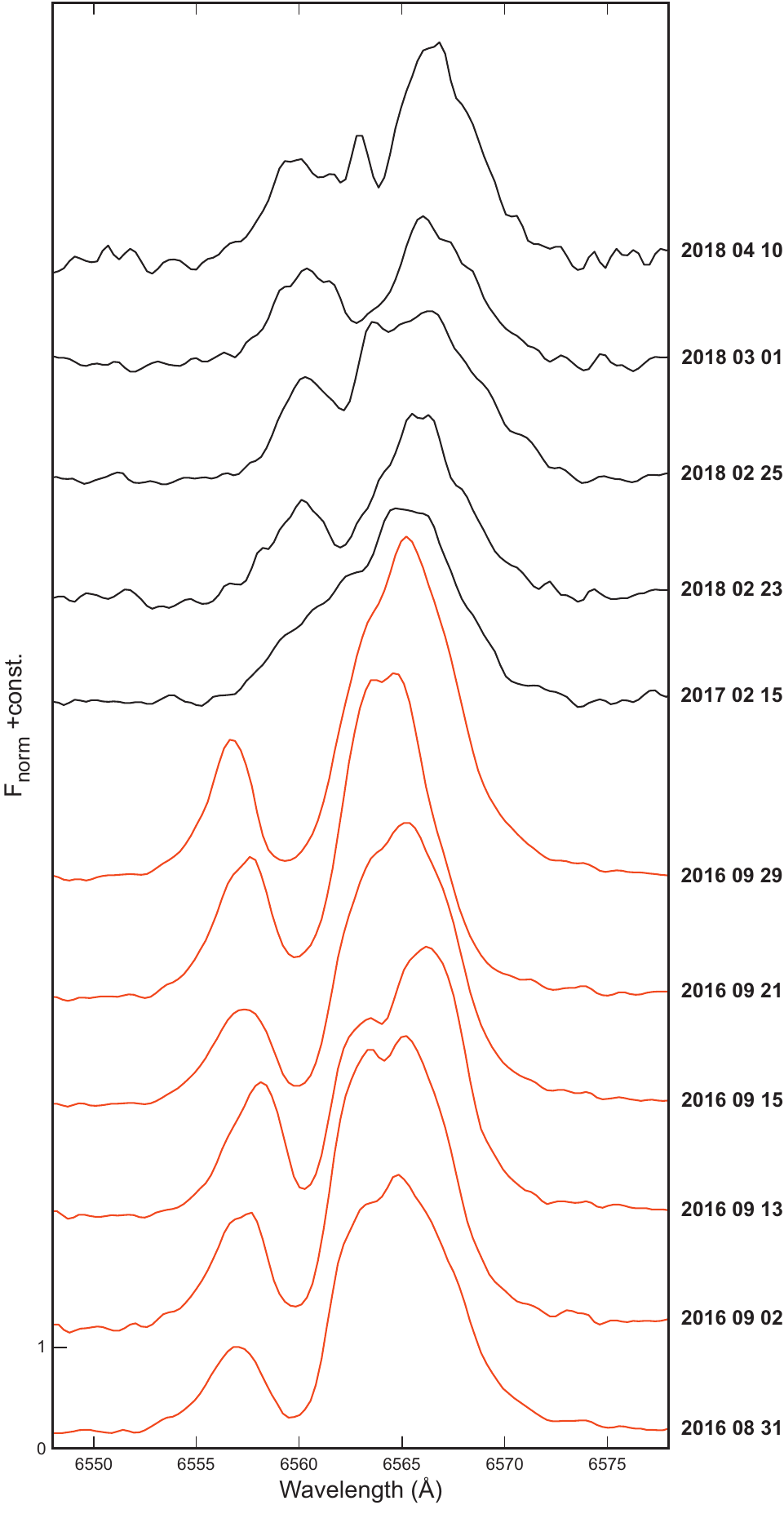} \includegraphics[height=5cm]{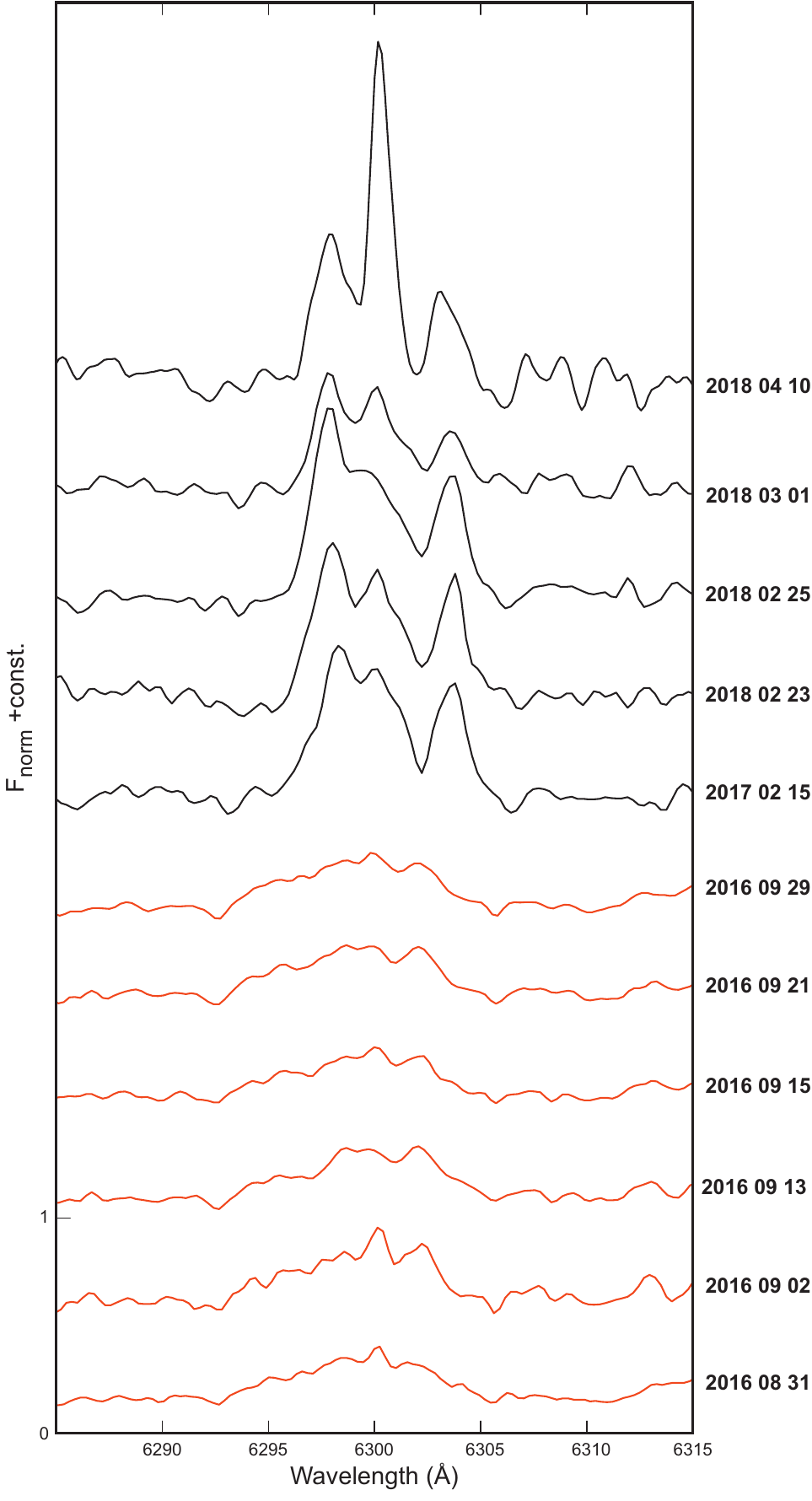}}
\caption{Sections of the normalized FLECHAS spectra of RW\,Aur from eleven observing epochs between 31 August 2016 and 10 April 2018, centered on the $\rm{H}\alpha$ emission line at 6563\,\AA~(left plot), and the $[\rm{O}I]$ emission line at 6300\,\AA~(right plot). Spectra that were taken during the bright state, after the second deep eclipse, are colored in red. The other spectra were taken during faint states and are shown in black.}
\label{spectra}
\end{figure*}

\begin{figure}
\resizebox{\hsize}{!}{\includegraphics{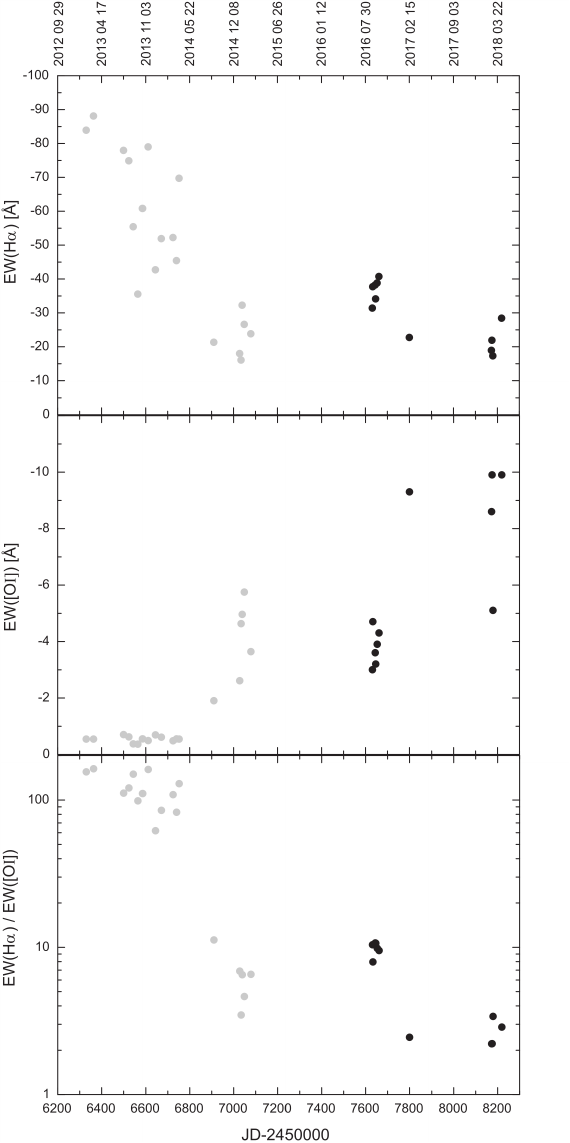}}
\caption{The equivalent width (EW) of the $\rm{H}\alpha$ and $[\rm{O}I]$ (6300\,\AA) emission line of RW\,Aur, and their ratio, plotted for spectra taken between February 2013 and April 2018. EWs from our campaign are shown with black circles, those from \citet{Bozhinova2016} with grey circles, respectively.}
\label{EWs}
\end{figure}

We focus our spectral analysis on the $\rm{H}\alpha$ emission at about 6563\,\AA~and the forbidden $[\rm{O}I]$ emission at about 6300\,\AA. These emission lines are typical signs of accretion in classical TTS, indicating shocked material at the stellar surface and mass outflow, respectively \citep{Bozhinova2016,Mohanty2005}. The corresponding parts of the normalized spectra are illustrated for all spectroscopic ob\-ser\-ving epochs in Figure\,\ref{spectra}\hspace{-1.5mm}. It can be seen that between 31 August and 29 September 2016, when RW\,Aur was in a bright state, the $\rm{H}\alpha$ emission is stronger than at the later observing epochs. In contrast, for the $[\rm{O}I]$ emission we note the opposite effect. The equivalent widths (EWs) of the two emission lines, as measured in the FLECHAS spectra in all observing epochs of our spectroscopic monitoring of RW\,Aur, are shown in Figure\,\ref{EWs}\hspace{-1.5mm} together with data from \citet{Bozhinova2016}. The EWs were measured in the individual FLECHAS spectra with \textit{IRAF} and exhibit a typical uncertainty of 0.3\,m\AA~for the $\rm{H}\alpha$ emission line, and 0.1\,m\AA~for the $[\rm{O}I]$ emission line, respectively. The absolute value of the EW of the $\rm{H}\alpha$ emission line of RW\,Aur is well above 10\,\AA, as expected according to the division between classical TTSs and weak-line TTSs \citep{Mohanty2005}. Our results are well consistent with the findings from \citet{Bozhinova2016}: When RW\,Aur\,A is in its dim state, the $\rm{H}\alpha$ emission, which originates from near the stellar surface, is partly obscured. At the same time, as our spectral observations cannot resolve RW\,Aur\,A and B, a larger fraction of the continuum emission comes from RW\,Aur\,B. This effect also reduces the $\rm{EW}(\rm{H}\alpha)$, because RW\,Aur\,B is a weak-line TTS, which exhibits a lower $\rm{H}\alpha$ emission. The $[\rm{O}I]$ emission, on the other hand, originates from mass outflow in the outer disk, which is not affected by the obscuration. Therefore, the decreasing continuum emission leads to an increase of the $\rm{EW}([\rm{O}I])$. However, compared to \citet{Bozhinova2016}, the $\rm{EW}(\rm{H}\alpha)/\rm{EW}([\rm{O}I])$ ratios are surprisingly low. The measurements by \citet{Bozhinova2016} until April 2014 were taken during bright state, before the second deep eclipse. Our measurements in August and September 2016 were taken when the apparent brightness had just recovered from the second deep eclipse, but $\rm{EW}(\rm{H}\alpha)/\rm{EW}([\rm{O}I])$ was still lower by an order of magnitude compared to the first observations by \citet{Bozhinova2016}. In particular, the $[\rm{O}I]$ emission remained almost as strong as during the obscuration and got even stronger when the star faded again towards the end of our observing campaign, indicating a higher mass outflow activity during this time.

Another notable feature in the spectra of RW\,Aur, is the prominent P\,Cygni profile of its $\rm{H}\alpha$ emission line, which is a further indicator for wind outflow. The absorption feature is blue-shifted with respect to the peak of the $\rm{H}\alpha$ emission line by up to 6\,\AA~(which corresponds to a wind velocity of about 275\,km/s) and was stronger at the beginning of our observing campaign, when the star was in its bright state. \citet{Bozhinova2016} also recorded this absorbing outflow and attributed it to the reason for the dimming. We can verify their finding that during the faint state, the absorbing material suppresses the blue-shifted $\rm{H}\alpha$ peak. The degree of absorption and the wind velocity are very variable. However, we note that the outflow also exists during the bright state, so it is not directly related to the degree of occultation. We also note that the blue-shifted peak of the P\,Cygni profile is much weaker compared to the main peak, whereas \citet{Bozhinova2016} found both peaks having similar strengths during the bright state. The main reason for this difference is an increased velocity and hence a stronger blueshift of the wind outflow: In the bright state, we get velocities of about 230\,-\,275\,km/s, while \citet{Bozhinova2016} only obtained 100\,-\,200\,km/s. Therefore, the wind now absorbs a larger portion of the blue-shifted peak.

Besides $\rm{H}\alpha$ and $[\rm{O}I]$, the spectra show several other emission lines with strong variations between bright and faint state. Our fully reduced and normalized spectra will be made available for download in the VizieR\footnote{Online available at: \url{https://vizier.cds.unistra.fr/viz-bin/VizieR}} database and can be used for more detailed studies.

\citet{Lisse2022} performed observations with SpeX at the NASA Infrared Telescope Facility \citep{Rayner2003} to provide spectra in the infrared (0.7\,-\,5.0\,$\mu$m). They show the evolution of four Hydrogen emission lines of the Paschen series between November 2018 and October 2020. The Paschen lines only show a single peak and, as for $\rm{H}\alpha$, they are very variable, being stronger in the brighter states. This is in contrast to the continuum behaviour in the infrared.

\section{Conclusions} \label{sec5}

We monitored the photometry of RW\,Aur in the B-, V-, R-, and I-band between September 2016 and April 2019, corresponding to the time after the second deep eclipse, when it went through strong brightness variations on timescales of up to one year, ending with a bright state in 2019. The observations were supported by spectra taken between August 2016 and April 2018, ending during a faint state.

The light curve fits to the photometric trend given by the AAVSO and other authors. The CMDs show that the reddening cannot be explained by interstellar extinction, but by large dust grains along the line of sight, related to grey extinction. This supports the scenario of a hot dusty wind emerging from the inner disk \citep{Petrov2015, Shenavrin2015, Bozhinova2016, Dodin2019}.

In our spectral analysis, we focused on the $\rm{H}\alpha$ and the $[\rm{O}I]$ (6300\,\AA) emission lines. Our results are largely consistent with \citet{Bozhinova2016} in terms of an increased $\rm{H}\alpha$ EW during bright state and an increased $[\rm{O}I]$ EW during faint state. However, in both states we obtain $\rm{EW}(\rm{H}\alpha)/\rm{EW}([\rm{O}I])$ ratios, which are much lower than before the start of the second deep eclipse in 2014. This shows that the re-brightening in 2016 \citep{Scholz2016} was not equivalent to a return of the system to its normal bright state. Instead, the outflow causing the $[\rm{O}I]$ emission was probably even stronger during the variable phases that occurred between 2016 and 2018, compared to the observing epoch between 2014 and 2015, reported by \citet{Bozhinova2016}. This is no contradiction, because the $[\rm{O}I]$ emission is only an indicator for the accretion-induced outflow activity but not for the amount of dust carried by the wind. Other reasons like the geometry of the system can also play a role when explaining the repeated brightening and dimming events during the time of our monitoring campaign.

RW\,Aur is an extraordinary, highly active system and a full understanding of its variability on different timescales and in different wavebands requires a very broad understanding of the different physical processes taking place in this young stellar environment. In particular, it requires a broad set of astronomical instruments and techniques to access all these phenomena. Large progress has been made in recent years, especially since the onset of the deep eclipses in between 2010 and 2011 as well as between 2014 and 2016. While it is now consent that the obscurations are caused by hot dust from the inner disk, the exact mechanisms which trigger the dust enhancement and transport it into the line-of-sight, are still debated. It is well possible that a combination of several proposed mechanisms is needed to explain the observed behavior. Beginning with the tidal encounter $\sim$500 years ago \citep{Dai2015} that might have caused a recent rearrangement or warping of the inner disk \citep{Berdnikov2017, Facchini2016, Koutoulaki2019}, over the formation, migration and destruction of planetesimals \citep{Guenther2018, Lisse2022}, which was suggested to explain the iron overabundance in X-ray spectra, up to the dust-laden wind that is now observed, inferred from the occultations and the IR excess \citep{Petrov2015, Shenavrin2015, Bozhinova2016, Dodin2019}.

Ongoing research on the $\sim$3\,Myr young RW\,Aur system will shed new light on the story of star- and planet formation. While it is now in a relatively constant, bright state, it will be interesting to see if there will be deep eclipses again in the future and if they will show periodicity. Further observations will make it possible to clarify the reasons for such dimming events and to unravel the secrets of the spectro-photometric variability of RW\,Aur and its role within the population of T Tauri stars.

\bibliography{RWAur}

\section*{Acknowledgments}

We want to thank the scientific and technical staff of the Astrophysical Institute of the Friedrich-Schiller-University Jena for their help and assistance in the observations. In particular, we want to thank the following observers: Alexoudi X., Andreas C., Bl\"{u}mcke M., Dadalauri M., Darie J., Geymeier M., Gilbert H., Hartung L., Heyne T., Hildebrandt F., Hoffmann S., Kr\"{u}ger E., Michel K.-U., Munz V., Pannicke A., Pyliagin D., Schiefeneder F., Schulz M., Stenglein W., Trepanovski A., Wagner D., Wischer M., Zehe T., Zielinski P. We made use of data from the \verb"Simbad" and \verb"VizieR" databases, operated at CDS in Strasbourg, France. Furthermore, we used data from the European Space Agency (ESA) mission \textit{Gaia} (\url{https://www.cosmos.esa.int/gaia}), processed by the Gaia Data Processing and Analysis Consortium (DPAC, \url{https://www.cosmos.esa.int/web/gaia/dpac/consortium}). Funding for the DPAC has been provided by national institutions, in particular the institutions participating in the Gaia Multilateral Agreement. This work made use of the AAVSO database, so we want to thank the many observers who provided their data to the AAVSO, as well as the organizers for collecting the data and providing tools like the light curve generator. Finally, we want to thank Ralph Neuh\"{a}user and Torsten L\"{o}hne for their helpful comments as well as Aleks Scholz, who drew the attention of the authors to RW\,Aur and inspired them to follow up on it.

\section*{Author Biography}

Oliver Lux obtained a Bachelor of Science in Physics and a Master of Science in astrophysics at the Universities of Bochum and Bonn, respectively, gaining knowledge about neutron stars. After that, he obtained a PhD at the University of Jena with a work on runaway stars and supernova remnants. His further research interests are variable stars and exoplanets.
\appendix\white{\section{}}

\begin{table*}
{\large{\textbf{APPENDIX A: PHOTOMETRY}}}
\vspace{1cm}
\caption{The apparent photometry of RW\,Aur during our photometric monitoring campaign. We list the Gregorian and Julian date ($JD$) of the photometric observations of the star together with the measured apparent B-, V- R-, and I-band magnitudes. In each night, the observations in the different photometric bands were performed consecutively, limiting the temporal uncertainties to only $\sim0.006$\,days.}
 \centering
\begin{tabular}{c|cccccc}
\hline
Epoch        & Date & $JD$ & B [mag] & V [mag] & R [mag] & I [mag]  \\
\hline
I  & 2016/09/07 & 2457638.556 & $11.954\pm0.012$ & $11.164\pm0.011$ & $10.557\pm0.010  $ & $~~9.789\pm0.009$ \\
   & 2016/09/13 & 2457644.550 & $11.155\pm0.011$ & $10.490\pm0.011$ & $ ~~9.974\pm0.009$ & $~~9.307\pm0.009$ \\
   & 2016/09/15 & 2457646.563 & $11.173\pm0.011$ & $10.494\pm0.011$ & $10.014\pm0.009  $ & $~~9.389\pm0.009$ \\
   & 2016/09/22 & 2457653.535 & $11.593\pm0.012$ & $10.952\pm0.011$ & $10.413\pm0.010  $ & $~~9.729\pm0.009$ \\
   & 2016/09/29 & 2457661.494 & $11.179\pm0.011$ & $10.585\pm0.011$ & $10.082\pm0.009  $ & $~~9.411\pm0.009$ \\
   & 2016/12/05 & 2457728.380 & $12.218\pm0.012$ & $11.582\pm0.011$ & $11.058\pm0.010  $ & $10.357\pm0.009 $ \\
   & 2016/12/20 & 2457743.323 & $12.696\pm0.013$ & $12.033\pm0.012$ & $11.424\pm0.010  $ & $10.656\pm0.010 $ \\
   & 2016/12/21 & 2457744.493 & $12.538\pm0.012$ & $11.879\pm0.012$ & $11.320\pm0.010  $ & $10.582\pm0.010 $ \\
   & 2017/01/06 & 2457759.533 & $12.794\pm0.013$ & $12.002\pm0.012$ & $11.346\pm0.010  $ & $10.523\pm0.010 $ \\
   & 2017/01/19 & 2457773.303 & $12.137\pm0.012$ & $11.434\pm0.011$ & $10.871\pm0.010  $ & $10.144\pm0.009 $ \\
   & 2017/01/28 & 2457782.461 & $13.011\pm0.013$ & $12.161\pm0.012$ & $11.487\pm0.010  $ & $10.648\pm0.010 $ \\
   & 2017/02/01 & 2457786.416 & $12.744\pm0.013$ & $11.932\pm0.012$ & $11.311\pm0.010  $ & $10.519\pm0.009 $ \\
   & 2017/02/13 & 2457798.333 & $13.280\pm0.015$ & $12.288\pm0.012$ & $11.561\pm0.010  $ & $10.719\pm0.010 $ \\
   & 2017/02/14 & 2457799.342 & $13.468\pm0.014$ & $12.432\pm0.012$ & $11.693\pm0.010  $ & $10.810\pm0.010 $ \\
   & 2017/02/15 & 2457800.334 & $13.218\pm0.014$ & $12.330\pm0.012$ & $11.612\pm0.010  $ & $10.754\pm0.010 $ \\
   & 2017/03/13 & 2457826.410 & ---              & $11.688\pm0.013$ & $11.038\pm0.010  $ & $10.243\pm0.010 $ \\
   & 2017/03/27 & 2457840.367 & $12.598\pm0.013$ & $11.839\pm0.012$ & $11.223\pm0.010  $ & $10.441\pm0.009 $ \\
   & 2017/03/30 & 2457843.295 & $12.260\pm0.013$ & $11.509\pm0.012$ & $10.907\pm0.010  $ & $10.175\pm0.009 $ \\
   & 2017/04/20 & 2457864.324 & ---              & $12.156\pm0.013$ & $11.466\pm0.011  $ & $10.646\pm0.010 $ \\
\hline
II & 2017/09/05 & 2458001.635 & $12.287\pm0.008$ & $11.398\pm0.008$ & $10.770\pm0.004 $ & $~~9.960\pm0.008$ \\
   & 2017/09/12 & 2458008.558 & $11.665\pm0.008$ & $10.850\pm0.008$ & $10.304\pm0.004 $ & $~~9.564\pm0.008$ \\
   & 2017/09/19 & 2458015.639 & $12.194\pm0.008$ & $11.412\pm0.010$ & $10.853\pm0.004 $ & $10.075\pm0.009 $ \\
   & 2017/10/15 & 2458041.680 & $11.198\pm0.008$ & $10.491\pm0.008$ & $10.000\pm0.004 $ & $~~9.310\pm0.008$ \\
   & 2017/10/16 & 2458042.673 & $11.081\pm0.008$ & $10.379\pm0.008$ & $~~9.929\pm0.004$ & $~~9.264\pm0.008$ \\
   & 2017/11/28 & 2458086.285 & $12.399\pm0.008$ & $11.597\pm0.009$ & $10.987\pm0.005 $ & $10.194\pm0.008 $ \\
   & 2017/12/07 & 2458095.311 & $12.237\pm0.008$ & $11.488\pm0.009$ & $10.954\pm0.005 $ & $10.215\pm0.008 $ \\
   & 2017/12/12 & 2458100.279 & $12.561\pm0.008$ & $11.763\pm0.009$ & $11.184\pm0.005 $ & $10.410\pm0.008 $ \\
   & 2018/01/15 & 2458133.528 & ---              & $11.726\pm0.009$ & $11.124\pm0.005 $ & $10.320\pm0.008 $ \\
\hline
IIb& 2018/01/30 & 2458149.288 & $12.880\pm0.046$ & $12.055\pm0.032$ & $11.472\pm0.013 $ & $10.631\pm0.012 $ \\
   & 2018/02/23 & 2458173.352 & $13.150\pm0.029$ & $12.265\pm0.016$ & $11.596\pm0.012 $ & $10.757\pm0.010 $ \\
   & 2018/02/24 & 2458174.413 & $13.111\pm0.031$ & $12.240\pm0.017$ & $11.606\pm0.012 $ & $10.779\pm0.011 $ \\
   & 2018/02/25 & 2458175.278 & $13.189\pm0.029$ & $12.241\pm0.016$ & $11.557\pm0.011 $ & $10.720\pm0.010 $ \\
   & 2018/02/26 & 2458176.434 & $13.057\pm0.033$ & $12.337\pm0.018$ & $11.741\pm0.012 $ & $10.924\pm0.010 $ \\
   & 2018/02/27 & 2458177.302 & $13.183\pm0.030$ & $12.289\pm0.016$ & $11.618\pm0.012 $ & $10.761\pm0.010 $ \\
   & 2018/02/28 & 2458178.375 & $13.054\pm0.032$ & $12.168\pm0.017$ & $11.535\pm0.012 $ & $10.698\pm0.010 $ \\
   & 2018/03/01 & 2458179.401 & $12.949\pm0.040$ & $12.063\pm0.019$ & $11.452\pm0.013 $ & $10.653\pm0.010 $ \\
   & 2018/03/24 & 2458202.328 & $13.309\pm0.037$ & $12.283\pm0.018$ & $11.580\pm0.012 $ & $10.715\pm0.010 $ \\
   & 2018/03/29 & 2458207.395 & $13.531\pm0.045$ & $12.482\pm0.019$ & $11.719\pm0.012 $ & $10.818\pm0.010 $ \\
\hline
\end{tabular}
\end{table*}

\setcounter{table}{0}

\begin{table*}
\caption{Continued.}
\label{lightcurvedata2}
\centering
\begin{tabular}{c|cccccc}
\hline
Epoch & Date & $JD$ & $B$ [mag] & $V$ [mag] & $R$ [mag] & $I$ [mag]  \\
\hline
III & 2018/08/16 & 2458346.611 & ---              & $10.350\pm0.009$ & $~~9.816\pm0.007$ &  ---               \\
    & 2018/08/17 & 2458347.599 & $10.735\pm0.011$ & $10.117\pm0.009$ & $~~9.638\pm0.007$ &  $~~9.020\pm0.007$ \\
    & 2018/08/22 & 2458352.618 & $10.984\pm0.011$ & $10.348\pm0.009$ & $~~9.813\pm0.007$ &  $~~9.151\pm0.007$ \\
    & 2018/08/23 & 2458353.611 & $11.015\pm0.011$ & $10.354\pm0.009$ & $~~9.830\pm0.007$ &  $~~9.185\pm0.007$ \\
    & 2018/09/06 & 2458367.606 & $10.793\pm0.011$ & $10.145\pm0.009$ & $~~9.629\pm0.007$ &  $~~8.982\pm0.007$ \\
    & 2018/09/08 & 2458369.595 & $10.860\pm0.011$ & $10.165\pm0.009$ & $~~9.631\pm0.007$ &  $~~9.013\pm0.007$ \\
    & 2018/09/17 & 2458378.627 & $10.684\pm0.011$ & $10.051\pm0.009$ & $~~9.892\pm0.007$ &  $~~9.204\pm0.007$ \\
    & 2018/09/18 & 2458379.644 & $10.310\pm0.011$ & $ 9.731\pm0.009$ & $~~9.290\pm0.007$ &  $~~8.776\pm0.007$ \\
    & 2018/09/20 & 2458381.652 & $11.433\pm0.011$ & $10.733\pm0.009$ & $10.141\pm0.007 $ &  $~~9.388\pm0.007$ \\
    & 2018/09/21 & 2458382.632 & $11.581\pm0.012$ &  ---             &  ---              & ---                \\
    & 2018/09/22 & 2458383.634 & $11.079\pm0.011$ & $10.437\pm0.009$ & $~~9.916\pm0.007$ &  $~~9.228\pm0.007$ \\
    & 2018/09/25 & 2458386.554 & $11.447\pm0.011$ & $10.819\pm0.009$ & $10.272\pm0.007 $ &  $~~9.550\pm0.007$ \\
    & 2018/09/28 & 2458389.632 & $11.301\pm0.011$ & $10.648\pm0.009$ & $10.098\pm0.007 $ &  $~~9.392\pm0.007$ \\
    & 2018/09/29 & 2458390.523 & $11.003\pm0.011$ & $10.366\pm0.009$ & $~~9.834\pm0.007$ &  $~~9.231\pm0.007$ \\
    & 2018/09/30 & 2458391.534 & $11.014\pm0.011$ & $10.392\pm0.009$ & $~~9.876\pm0.007$ &  $~~9.222\pm0.007$ \\
    & 2018/10/01 & 2458392.616 & ---              & $10.173\pm0.009$ & $~~9.656\pm0.007$ &  $~~9.037\pm0.007$ \\
    & 2018/10/06 & 2458397.620 & $11.534\pm0.011$ & $10.824\pm0.009$ & $10.228\pm0.007 $ &  $~~9.481\pm0.007$ \\
    & 2018/10/10 & 2458401.681 & $10.942\pm0.011$ & $10.274\pm0.009$ & $~~9.760\pm0.007$ &  $~~9.143\pm0.007$ \\
    & 2018/10/11 & 2458402.677 & $10.978\pm0.011$ & $10.392\pm0.009$ & $~~9.887\pm0.007$ &  $~~9.212\pm0.007$ \\
    & 2018/10/13 & 2458405.428 & $11.178\pm0.011$ & $10.529\pm0.009$ & $~~9.989\pm0.007$ &  $~~9.341\pm0.007$ \\
    & 2018/10/14 & 2458406.470 & $11.431\pm0.011$ & $10.731\pm0.009$ & $10.193\pm0.007 $ &  $~~9.535\pm0.007$ \\
    & 2018/10/17 & 2458409.410 & $11.384\pm0.011$ & $10.643\pm0.009$ & $10.067\pm0.007 $ &  $~~9.415\pm0.007$ \\
    & 2018/10/21 & 2458413.358 & $11.005\pm0.011$ & $10.390\pm0.009$ & $~~9.848\pm0.007$ &  $~~9.191\pm0.007$ \\
    & 2018/10/23 & 2458414.552 & $10.985\pm0.011$ & $10.293\pm0.009$ & $~~9.748\pm0.007$ &  $~~9.157\pm0.007$ \\
    & 2018/11/06 & 2458428.558 & $11.149\pm0.011$ & $10.504\pm0.009$ & $~~9.964\pm0.007$ &  $~~9.290\pm0.007$ \\
    & 2018/11/10 & 2458432.592 & $10.513\pm0.011$ & $ 9.892\pm0.009$ & $~~9.413\pm0.007$ &  $~~8.871\pm0.007$ \\
    & 2018/11/14 & 2458436.650 & $10.700\pm0.011$ & $10.067\pm0.009$ & $~~9.579\pm0.007$ &  $~~8.995\pm0.007$ \\
    & 2018/11/16 & 2458438.638 & ---              & $10.107\pm0.009$ & $~~9.619\pm0.007$ &  $~~9.002\pm0.007$ \\
    & 2018/11/17 & 2458439.573 & $10.738\pm0.011$ & $10.152\pm0.009$ & $~~9.681\pm0.007$ &  $~~9.053\pm0.007$ \\
    & 2018/11/18 & 2458440.516 & ---              & $10.186\pm0.009$ & $~~9.721\pm0.007$ &  $~~9.145\pm0.007$ \\
    & 2019/01/18 & 2458502.237 & $ 9.939\pm0.011$ & $ 9.388\pm0.009$ & $~~9.005\pm0.007$ &  $~~8.483\pm0.007$ \\
    & 2019/01/21 & 2458505.216 & $10.334\pm0.011$ & $ 9.730\pm0.009$ & $~~9.273\pm0.007$ &  $~~8.722\pm0.007$ \\
    & 2019/01/21 & 2458505.220 & $10.357\pm0.011$ & $ 9.760\pm0.009$ & $~~9.298\pm0.007$ &  $~~8.735\pm0.007$ \\
    & 2019/01/31 & 2458515.230 & $10.540\pm0.011$ & $ 9.931\pm0.009$ & $~~9.471\pm0.007$ &  $~~8.874\pm0.007$ \\
    & 2019/02/08 & 2458522.521 & ---              & $ 9.995\pm0.009$ & $~~9.548\pm0.007$ &  $~~8.984\pm0.007$ \\
    & 2019/02/14 & 2458529.251 & $11.674\pm0.011$ & $10.980\pm0.010$ & $10.351\pm0.007 $ &  $~~9.601\pm0.007$ \\
    & 2019/02/23 & 2458538.250 & $11.354\pm0.011$ & $10.656\pm0.009$ & $10.064\pm0.007 $ &  $~~9.320\pm0.007$ \\
    & 2019/02/24 & 2458539.255 & $10.746\pm0.011$ & $10.103\pm0.009$ & $~~9.573\pm0.007$ &  ---               \\
    & 2019/02/25 & 2458540.256 & $10.933\pm0.011$ & $10.247\pm0.009$ & $~~9.689\pm0.007$ &  $~~9.039\pm0.007$ \\
    & 2019/02/26 & 2458541.261 & $10.373\pm0.011$ & $ 9.779\pm0.009$ & $~~9.299\pm0.007$ &  $~~8.742\pm0.007$ \\
    & 2019/02/27 & 2458542.259 & $10.989\pm0.011$ & $10.382\pm0.009$ & $~~9.840\pm0.007$ &  $~~9.133\pm0.007$ \\
    & 2019/03/07 & 2458550.263 & $11.164\pm0.011$ & $10.532\pm0.009$ & $10.002\pm0.007 $ &  $~~9.275\pm0.007$ \\
    & 2019/03/20 & 2458563.272 & $11.123\pm0.011$ & $10.529\pm0.010$ & $10.002\pm0.007 $ &  $~~9.281\pm0.007$ \\
    & 2019/03/21 & 2458564.275 & $10.642\pm0.011$ & $ 9.998\pm0.009$ & $~~9.519\pm0.007$ &  $~~8.946\pm0.007$ \\
    & 2019/03/22 & 2458565.293 & $10.960\pm0.011$ & $10.303\pm0.009$ & $~~9.800\pm0.007$ &  $~~9.212\pm0.007$ \\
\hline
\end{tabular}
\end{table*}

\setcounter{table}{0}

\begin{table*}
\caption{Continued.}
\label{lightcurvedata3}
\centering
\begin{tabular}{c|cccccc}
\hline
Epoch             & Date & $JD$ & $B$ [mag] & $V$ [mag] & $R$ [mag] & $I$ [mag]  \\
\hline
III & 2019/03/23 & 2458566.302 & $11.276\pm0.012$ & $10.631\pm0.011$ & $10.066\pm0.008 $ & ---               \\
    & 2019/03/24 & 2458567.310 & $10.942\pm0.011$ & $10.293\pm0.010$ & $~~9.786\pm0.007$ & $~~9.184\pm$0.007 \\
    & 2019/03/29 & 2458572.286 & $10.733\pm0.011$ & $10.180\pm0.009$ & $~~9.722\pm0.007$ & $~~9.100\pm$0.007 \\
    & 2019/03/30 & 2458573.281 & $11.073\pm0.011$ & $10.406\pm0.010$ & $~~9.859\pm0.007$ & $~~9.207\pm$0.007 \\
    & 2019/03/31 & 2458574.335 & $11.277\pm0.011$ & ---              & $~~9.969\pm0.007$ & $~~9.256\pm$0.007 \\
    & 2019/04/01 & 2458575.293 & $10.848\pm0.011$ & $10.223\pm0.009$ & $~~9.719\pm0.007$ & $~~9.119\pm$0.007 \\
    & 2019/04/10 & 2458584.312 & $10.749\pm0.011$ & $10.201\pm0.009$ & $~~9.713\pm0.007$ & $~~9.037\pm$0.007 \\
    & 2019/04/15 & 2458589.304 & $11.132\pm0.012$ & $10.498\pm0.010$ & $~~9.958\pm0.007$ & $~~9.249\pm$0.007 \\
    & 2019/04/16 & 2458590.317 & $10.900\pm0.011$ & $10.279\pm0.010$ & $~~9.780\pm0.007$ & $~~9.139\pm$0.007 \\
    & 2019/04/17 & 2458591.318 & $10.744\pm0.011$ & $10.142\pm0.009$ & $~~9.674\pm0.007$ & $~~9.062\pm$0.007 \\
    & 2019/04/18 & 2458592.307 & $11.296\pm0.013$ & $10.608\pm0.011$ & $10.031\pm0.007 $ & $~~9.316\pm$0.007 \\
    & 2019/04/19 & 2458593.328 & $11.409\pm0.012$ & $10.749\pm0.010$ & $10.176\pm0.007 $ & $~~9.476\pm$0.007 \\
    & 2019/04/20 & 2458594.320 & $10.809\pm0.011$ & $10.213\pm0.009$ & $~~9.728\pm0.007$ & $~~9.104\pm$0.007 \\
    & 2019/04/21 & 2458595.346 & $10.968\pm0.011$ & $10.352\pm0.010$ & $~~9.844\pm0.007$ & $~~9.188\pm$0.007 \\
\hline
\end{tabular}
\label{photodata}
\end{table*}

\end{document}